\documentclass[usenatbib]{mn2e}
\usepackage{epsfig}
\usepackage{amsmath,amssymb}
\newcommand{\G}{{\rm G}}

\newcommand{\Msol}{{\,\rm M}_\odot}

\newcommand{\Gyr} {{\,\rm Gyr}}
\newcommand{\kpc} {{\,\rm kpc}}

\newcommand{\rej}{_{rej}}
\newcommand{\vir}{_{vir}}

\renewcommand{\vec}[1]{{\mathbf #1}}
\title{The concentration-velocity dispersion relation in galaxy groups}
\author[A. Faltenbacher and W.G. Mathews]{
{\parbox[t]\textwidth{Andreas Faltenbacher$^{1,2}$ 
and 
William G. Mathews$^2$}} 
\vspace*{6pt}\\
$1$ Shanghai Astronomical Observatory,
80 Nandan Road, Shanghai 200030, 
China\\
$2$ UCO/Lick Observatory,
University of California at Santa Cruz,
1156 High Street, Santa Cruz, CA 95064, USA
}
\date{\today}
\begin{document}
\maketitle
\begin{abstract}
Based on results from cold dark matter N-body simulations we
develop a dynamical model for the evolution of subhaloes within group
sized host haloes. Only subhaloes more massive than $5\times
10^8\Msol$ are considered, because they are massive enough to possibly
host luminous galaxies. On their orbits within a growing host potential the
subhaloes are subject to tidal stripping and dynamical friction.
At the present time ($z=0$) all model hosts have equal mass ($M\vir =
3.9 \times 10^{13} \Msol$) but different concentrations associated with
different formation times. We investigate the variation of subhaloe
(or satellite galaxy) velocity dispersion with host concentration
and/or formation time. In agreement with the Jeans equation the
velocity dispersion of subhaloes increases with the host
concentration.  Between concentrations $\sim5$ and $\sim20$ the
subhaloe velocity dispersions increase by factor of $\sim1.25$. By
applying a simplified tidal disruption criterion, 
i.e. rejection of all subhaloes with a tidal truncation radius below
$3\kpc$ at $z=0$, the central velocity dispersion of the 'surviving'
subhaloe sample increases substantially for all concentrations. The
enhanced central velocity dispersions in the surviving subhaloe
samples are caused by a lack slow tangential motions. 
Additionally, we present a fitting formula for the anisotropy
parameter which does not depend on concentration if the group-centric
distances are scaled by $r_s$, the characteristic radius of the
NFW-profile. Since the expected loss of subhaloes and galaxies due to 
tidal disruption increases the velocity dispersion of surviving
galaxies, the observed galaxy velocity dispersion can substantially
overestimate the virial mass.
\end{abstract}
\section{Introduction}
Clusters of galaxies have been extensively studied in all
available wave bands. They belong to the most massive and eye-catching
structures in the Universe and are thought to closely resemble the
cosmic composite of 85 per cent dark matter and 15 per cent baryonic
matter \citep{Spergel06}. According to the hierarchical model for
structure assembly, these systems preferentially have formed late. For
dark matter haloes it has been shown that the formation redshift is
correlated with concentration (e.g.~\citealt{Bullock01b,Wechsler02})
therefore clusters present a class of low concentration objects. In
contrast, groups of galaxies span a whole range of formation times and
concentrations. Some resemble the properties of clusters
\citep{Zabludoff98, Mulchaey00, Zabludoff00}, having low
concentrations, others show an opposite behaviour.

Also the properties of the galaxy populations vary strongly amongst
individual groups. Some groups have luminosity
functions similar to those observed in clusters but 
'fossil groups' display a central giant
elliptical galaxy with only few intermediate-luminosity
galaxies (e.g. \citealt{Ponman94, Jones2003}). The optical properties
of the central bright galaxies are consistent with an origin as the
merged combination of a more typical, less over-luminous, galaxies. 
The variations amongst the satellite populations are thought to
originate from varying formation times. Using N-body experiments
\cite{DOnghia05} demonstrated that the earliest formed group-sized dark
matter haloes reveal structural similarity to observed 
fossil groups (see also \citealt{SommerLarsen06}).
The early formation times of fossil groups also impacts
satellite galaxies that do not merge with the central giant. 
Based on the observed depletion of intermediate-luminosity galaxies in
the fossil group RX J1552.2+2013, \cite{MendesdeOliveira06} suggests that
the members of this system may be affected by tidal
disruption. \cite{DaRocha05} find substantial amount of intra-group
light in two of three compact groups which may, at least partly,
come from tidally dissolved satellite galaxies. These observations
indicate a substantial variation of the satellite population with
time either by merging or by tidal dissolution.

In addition to the concentration-age and the luminosity-age relations,
we describe here a concentration-velocity relation. According to the Jeans
equation, galaxies orbiting in more concentrated host haloes have higher
velocity dispersions. This effect is most pronounced in the central
regions of the host haloe. As discussed above, early
formed groups are more concentrated, leading to the conclusion that
the central satellite galaxies in fossil groups move faster than
central galaxies in less evolved groups of comparable
size. Additionally, if tidal disruption reduces the number of
satellite galaxies, the velocity dispersion of the remaining galaxies
is expected to increase even higher.  The last statement is supported by
the following consideration. Analysing N-body simulations,
\cite{Diemand04a} find an enhanced velocity dispersion of subhaloes 
compared to the dispersion of the diffuse dark matter
component. Subsequently, \cite{Faltenbacher06} argued that the loss of
a slow moving, earlier accreted, subset of the subhaloe population is
responsible for this bias. In analogy, if satellite galaxies are prone
to tidal dissolution, then early accreted, slow moving, galaxies are
preferentially dissolved. The lack of low velocity galaxies causes 
an enhanced velocity dispersion of the remaining galaxies. An
observational hint for such a scenario may be the recent observation
of the fossil system RX J1416.4+2315 by \citep{Khosroshahi06}. Based
on 18 member galaxies they find a velocity dispersion
that is nearly twice as high as expected from the X-ray analysis of
this system. 

These findings motivated us to study the relation between the host
concentration and dynamics of the subhaloe populations.  
The paper is organised as follows. \S~\ref{sec:model} introduces the
dynamical model used to trace the evolution of substructure in host
haloes of different mass accretion histories. \S~\ref{sec:point} deals
with substructure as point-like particles, a reference for
the subsequent more complex investigation. \S~\ref{sec:subhaloe}
presents dynamical results when substructures are treated as extended
units prone to tidal stripping and dynamical friction. Finally,
\S~\ref{sec:summary} concludes with a summary. 
\section{Dynamical model}	  
\label{sec:model}
We aim to investigate the correlation between the host concentration 
and subhaloe dynamics. In principle cosmological N-body simulations
can best tackle these kinds of problems. However, such
simulations should cover a large volume for satisfactory host haloe
statistics and, at the same time, provide high resolution for the
accurate determination of the subhaloe dynamics. Here we pursue an
alternative approach. We employ semi-analytical models to gain a
basic understanding of the processes which shape the subhaloe
distributions within equal mass hosts of various concentrations. 

In recent years high resolution CDM simulations have substantially
improved our understanding of cosmic structure formation. This insight
can be used to model the dynamical evolution of subhaloe in a
semi-analytical manner.  Recently, \cite{Zentner05} demonstrated that
this kind of semi-analytical modelling achieves good agreement with
state of the art N-body simulations. Similar approaches have been used
for a variety of scientific goals (see e.g.~\citealt{Bullock00, Bullock01,
Zentner03, Koushiappas04, Taylor01, Taylor04, Islam03,
VanDenBosch05}).

Here we extend the models presented in \cite{Mathews04} and
\cite{Faltenbacher05}. Originally, these models were designed to
reproduce the well observed number density distribution  
of satellite galaxies of the NGC 5044 group. These models have been
modified for the purposes of present investigation. 
Basically, the model integrates the orbits of subhaloes from the
accretion onto the host to the present ($z=0$) within the deepening
gravitational potential of the host haloe. By means of this the origin of
the present subhaloe distribution can be investigated. Also, the code
allows us to assign a mass profile to the entering subhaloes which in
turn can be used to compute the modification of their orbits by tidal mass
loss and dynamical friction.
\subsection{Host haloe}
Based on the analysis of N-body simulations
\cite{Wechsler02} found that the mass growth of CDM haloes can be
described by
\begin{equation}
\label{equ:evoma}
M_v(a) = M_{v,0}e^{-2a_f\left({1\over a}-1\right)}
\end{equation}
where $a = 1/(1+z)$ and $a_f = 1/(1+z_f)$ is the cosmic expansion
factor at the formation redshift $z_f$. Given the present
virial mass of a haloe, the mass accretion history is fully determined
by the only remaining free parameter $a_f$. Individual accretion
histories may differ strongly from this description, but on average
the mass accretion is well approximated by this formula. Most
importantly, the mass built-up of a haloe is separated into two distinct
phases, namely an early rapid accretion phase and a subsequent period
of modest accretion. Knowing $M_v(a)$ also allows to compute the
evolution of the virial radius $R\vir(a)$. According to the spherical
collapse model the virial radius $R\vir(a)$ is the radius which includes
a mean density of $\Delta(a)\rho_c(a)$, where $\Delta(a)$ is the
virial over-density in units of the critical density $\rho_c(a)$ (see
e.g.~\citealt{Eke98, Bryan98}).    
\begin{equation}
\label{equ:evorv}
R\vir = \left({3M(a)\over4\pi}{1\over\Delta(a)\rho_c(a)}\right)^{1/3} 
\end{equation}
Anticipating an NFW density profile \citep{NFW97}, the host
mass and concentration are sufficient to completely determine its
properties. According to \cite{Wechsler02} 
the evolution of the concentration is given by
\begin{equation}
\label{equ:evoco}	
c\vir = {c_1 a_0 \over a_f}
\end{equation}
where $c_1=5.125$ according to \cite{Zentner05} and
$a_0$, the cosmic expansion factor at the time of observation, equals
1 in this context. In our model Eqs.~\ref{equ:evoma} and
~\ref{equ:evoco} completely characterise the growth of the  host and
its gravitational potential which in turn determines the orbits  of
the subhaloes. At $z=0$ all host haloes approach the same virial mass
($M_{v,0} = 3.9\times 10^{13}\Msol$), but they exhibit concentrations
in the range between 2 and 20 (integer steps, resulting in 19 output
sets). The low concentration haloes are currently in the rapid
accretion phase. According to Eq.~\ref{equ:evoco} the formation of  
the least concentrated haloes takes place in future times. The most
concentrated haloes formed at redshifts $z\sim3$.     
The evolution of the host haloe as described here does not explicitly
include merging events. Such events may rearrange the phase space
distributions of the subhaloes. However, the rapid accretion phase at
early times causes the fastest change of the potential over the whole
accretion history of a haloe. To a certain extent these violent
merging processes are covered by the steep slope of the accretion
(equation ~\ref{equ:evoma}) at $z_f$.

In summary, the mass growth of the host haloes modelled according to
Eq.~\ref{equ:evoma} provides the background potential for the
dynamical evolution of the subhaloes. All hosts have a mass of
$M_{v,0} = 3.9\times 10^{13}\Msol$ at $z=0$ and concentrations ranging
from 2 to 20 (integer steps). According to Eq.\ref{equ:evoco}
concentrations  are associated with formation times, resulting in
different accretion histories for the host haloes despite their
equal masses. 
\subsection{Subhaloes}
After introducing the model for the host we now descirbe 
the host's substructure population. Two samples of 100000 orbits are
traced for every host haloe. The first sample, referred to as the {\it
point mass sample}, follows the evolution of point-like particles
within the potential of the host. In the second sample, dubbed the
{\it subhaloe sample}, the orbits are 
modified by the effects of tidal stripping and dynamical friction
which involve the attribution of density profiles to each
substructure. The point mass sample is mainly  used to test the agreement of
the model with N-body simulations on a particle basis. The subhaloes 
in the subhaloe sample have masses $\gtrsim 5\times10^8\Msol$ which
renders them possible candidates for hosting galaxies. Obviously, the
sum of all 100000 subhaloe masses exceeds the mass of the host
haloe. Therefore, the results for the subhaloe distributions should be
considered as the outcome of stacking many identical groups. In the
next two paragraphs we describe the generation of the initial
conditions and initial properties of the subhaloes in the subhaloe
sample.  
\subsubsection{Orbital initial conditions}
We assume that the initial space density of the subhaloes is proportional
to the dark matter, i.e. subhaloes enter the host haloe in proportion to
the dark matter. To generate this particular accretion pattern 
the individual arrival times of the subhaloes are assigned by a random
process using the mass of the host as a proxy for the cosmic time.  
\begin{equation}
\label{equ:mira}
M_i = \mathcal{R}_A M_{v,0} 
\end{equation}
Here $0\leq\mathcal{R}_A\leq1$ is a random number and $M_{v,0}$ the
virial mass of the host at $z=0$. By setting $M_i= M(a)$
(Eq.~\ref{equ:evoma}) $M_i$ can be converted into the cosmic expansion
factor $a_i$ or equivalently $z_i$, which is considered
as the accretion time of the $ith$ subhaloe. Subsequently
Eq.~\ref{equ:evorv} is used to derive virial radius of the host,
$r_i$, at that time. The $ith$ subhaloe is assumed to be
gravitationally accelerated by $M_i$ from standstill at the turnaround
radius ($r_{turn} = 2r_i$) until it reaches the virial radius, $r_i$. 
The resulting absolute values of the velocity at $r_i$ is 
\begin{equation}
\label{equ:vinit}
u_i= \left(\G M_i/r_i\right)^{1/2}
\end{equation}
which also equals the circular velocity at $r_i$. To assign an angular
momentum to each subhaloe we invoke another random process designed to
reproduce the distribution of circularities found in N-body
simulations. The circularity is defined as $\epsilon \equiv J/J_c$
where $J$ is the angular momentum of the orbit and and $J_c$ is the
angular momentum of a circular orbit with the same 
orbital energy (see e.g. \citealt{Lacey93,Tormen97}).
\cite{Zentner05} present the orbital circularity
distribution for subhaloes in N-body simulations at the time of
entering the virial radius of the host. They also provide a fitting
formula for the distribution and state that it is independent of
accretion redshift and subhaloe mass. Here we model the probability
distribution of circularities with a simple quadratic, $dp/d\epsilon =
6\epsilon(1-\epsilon)$, which is in fairly good agreement with the
data presented by \cite{Zentner05} and has a mean circularity of
$\langle\epsilon\rangle=0.5$. The initial angular momentum of 
a subhaloe is randomly drawn from this distribution, 
\begin{equation}
\mathcal{R}_L = \int_0^\epsilon 6\epsilon(1-\epsilon)d\epsilon =
3\epsilon^2 - 2\epsilon^3\ , 
\end{equation} 
where $\mathcal{R}_L$ is a random number between zero and one.
Note, the distributions of circularities were derived for subhaloes in 
N-body simulations,  here we implicitly assume that the same
regularity holds for the point mass sample as well. 

Knowing the evolution of the host potential, the accretion
time and the initial orbital parameters of each subhaloe, the
integration of the orbits can be pursued. Details about the
integration are presented at the end of the following paragraph. Two 
random processes are involved to generate the orbital initial
conditions, namely $\mathcal{R}_A$ for the arrival time and 
$\mathcal{R}_L$ for the initial angular momentum. We assume, that the
orbital energy is uniquely determined by the host virial mass and
radius at the time of arrival, which may not be exactly the case (see
\citealt{Zentner05}). However, for simplicity we do not implement
an additional mechanism to spread the distribution of initial orbital
energies. 
\subsubsection{Subhaloe properties}
\label{sec:subprop}
As long as the subhaloes are treated as point-like particles and
dynamical friction is ignored, the orbits are independent 
of the subhaloe masses. For a more realistic model we must
consider the spatial extension of the subhaloes and as a consequence
the susceptibility to tidal truncation. For that purpose a
mass is attributed to every arriving subhaloe by a random process
which is set up to to generate a power law mass function as observed
for the field haloes in N-body simulations. We adopt $dM/dN =
M^\alpha$ where $\alpha = -1.86$
\citep{Somerville99,Reed03}. Consequently, the mass of the $ith$ haloe
$m_i$ at arrival is given by  
\begin{equation}
\label{equ:mafu}
m_i =  \left[M_1^{x}-\mathcal{R}_M(M_1^{x}-M_2^{x})\right]^{1/x}
\end{equation}
where ${R}_M$ is a random number between 0 and 1, $x=1+\alpha$. The
lower mass limit is $M_1 = 5\times10^8\Msol$ and the upper mass limit
is $M_2=M_i$, where $M_i$ is the current host mass
(see Eq.\ref{equ:mira}). We only focus on the evolution of dark 
matter haloes above $M_1 = 5\times10^8\Msol$ because they are massive
enough to potentially host visible galaxies. The upper mass limit is
chosen to avoid merging of haloes more massive than the current host.
Note, since the mass function is so steep our results do not depend on
the exact choice of the upper mass limit. 
This restriction assures that we follow the host merging tree along
the most massive progenitors. Ignoring the possible deviations caused
by the centrally condensed baryonic matter, we associate an NFW density 
profile \citep{NFW97} to every arriving subhaloe. We assume that the
formation redshift of all (sub)haloes is $z_{sub,f}=7$ (which corresponds
to $a_{sub,f} = 0.125$). By setting the arrival time $a_i = a_0$ in
Eq.~\ref{equ:evoco} the concentration $c_i$ of the subhaloe is
computed. The density profile of the subhaloe is
completely determined by $m_i$ and $c_i$. We assume that on the
subsequent orbits the 
shape of the density profile does not change (see
\citealt{Kazantzidis04}) however it can be truncated due to tidal
forces.  The tidal truncation radius is estimated by the Jacobi limit
(see \citealt{Binney87,Hayashi04}),  
\begin{equation}
\label{equ:jacobi}
r_J = \left({m_{c,i}\over3M(<D)}\right)^{1/3}D
\end{equation}
Here $D$ is the instantaneous distance of the subhaloe from the centre
of the host, $M(<D)$ is the host mass within $D$ and $m_{c,i}$ is the 
current mass of the subhaloe. If $r_J$ is smaller than the current
radius of the subhaloe, all the matter outside of $r_J$ is assumed to
be stripped instantaneously and the current haloe radius is replaced
within $r_J$.

The orbits of the subhaloes are found by solving 
\begin{equation}
{d\vec{r}\over dt} = \vec{u} 
\hspace{0.1\hsize}\text{and}\hspace{0.1\hsize}
{d\vec{u}\over dt} = - {\G M(r)\over r^2}{\vec{r}\over r} +
\left({d\vec{u}\over dt}\right)_{df} 
\end{equation} 
where $M(r)$ is the NFW mass within radius $r$ of the host. Orbits
are computed from redshift $z_i$ (redshift at arrival) 
to the present epoch, $z=0$. For the integration we use a fourth order
Runge-Kutta integration scheme with adaptive time stepping.

The deceleration by dynamical friction 
(\citealt{Chandrasekhar43, Colpi99}) is described by 
\begin{equation}
\label{equ:chan}
\left({d\vec{u}\over dt}\right)_{df} = 
-\vec{u} 4 \pi \ln\Lambda \G^2 m \rho
u^{-3}[{\rm erf}(X)-{2\over\pi^{1/2}}Xe^{-X^2}]\ . 
\end{equation}
Here $\rho(r,t)$ is the local density of the host haloe, 
$u=|\vec{u}|$ is the velocity of the subhaloe, $X =
u/(\sqrt{2}\sigma)$ and $\sigma(r,t)$ is the mean velocity
dispersion of diffuse dark matter in the host halo. We assume that the
total dispersion can be approximated   
by the cold dark matter dispersion presented by \cite{Hoeft04}
(see also \citealt{Mathews04}). We choose $\ln\Lambda = 3$ as
suggested by \cite{Zhang02} which is the point mass
approximation. This approximation is justified by the very efficient 
tidal mass stripping which rapidly truncates the radius of the subhaloe
to a few per cent of the virial radius of the host.   

In summary, a mass is randomly assigned to each arriving subhaloe. 
The random process is designed to recover the mass functions of field
haloes found in N-body simulations. Additionally a concentration is
assigned to the subhaloe assuming a common (sub)haloe formation redshift
of $z=7$. Based on mass and concentration an NFW density profiles is
associated with each subhaloe. The profile is assumed to maintain its
shape on the subsequent orbit within the host potential well, however
tidal truncation is allowed to take place. The subhaloe gets most
severely truncated at its peri-centre passage. The remaining mass
remains constant until the subhaloe approaches the centre for the next
time. Between two subsequent peri-centre passages the subhaloe loses
some orbital energy due to dynamical friction, the peri-centre
distance shrinks with each passage, reducing the truncation radius. Of
course, lower masses experience less dynamical 
friction. The mass of the subhaloes decreases in a step-like 
fashion, as also observed in N-body simulations
(e.g.\citealt{Kravtsov04, BoylanKolchin06}).
     
In order to investigate how tidal reduction affects the properties 
of the subhaloe population we will construct a sub-subhaloe sample
containing only subhaloes having present truncation radii above the
{\it rejection} radius $r\rej\gtrsim3\kpc$. This choice is
somewhat arbitrary, but in agreement with values given in
\cite{Faltenbacher05} who achieve good agreement between the outcome
of a very similar dynamical model for the observed galaxy group NGC
5044. Subsequently, we refer to the two subsamples as then {\it complete}
and {\it reduced} samples, emphasising that the latter is a
subset of the former. Note, disappearance when $r\leq r\rej$ in the
reduced subhaloe sample does not necessarily mean
total physical disassembly of the object, it may just mean that it
falls below a given detection limit. Such limits are apparent in
both simulations and observations. The detection of subhaloes in
simulations is constricted by a minimum particle number limit for
substructures and the detection of satellite galaxies in groups
depends on the sensitivity of the observations.
\section{Results for the point mass sample}	
\label{sec:point}
The point mass sample follows the evolution of point-like particles
within the growing potential well of equal-mass host haloes with
varying accretion histories. Mass dependent mechanisms, like tides and
dynamical
friction, are excluded.  The mass of the particles is irrelevant. In
this section we (1) demonstrate the agreement of the model with
established results from N-body simulations and (2) discuss the
concentration-sigma relation for this simplified approach as a
reference for the results for the (complete and reduced) subhaloe
samples presented in the following section.    
\subsection{Accretion histories}
\begin{figure}
\epsfig{file=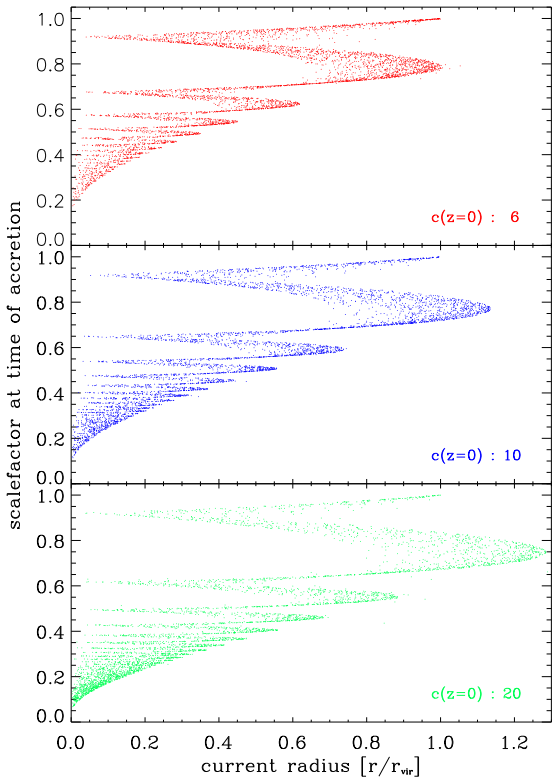,width=0.95\hsize}
\caption{\label{fig:radac}
Scale factor at time of accretion versus current subhaloe distance to
the centre for the three example host haloe concentrations of 6, 10
and 20 at $z=0$. For clarity, only 5 per cent of the 100000 particles,
randomly selected from the complete  samples, are displayed. Even if
weakened due to the imposed angular momentum distribution, there is a
clear indication for the appearance of caustics as described in
secondary infall model by Bertschinger (1985). The caustic of the
second turnaround is located at or beyond the current virial radius.  
The caustic of the most concentrated host extends furthermost, 
indicating the highest orbital energies.   
}
\end{figure}
\begin{figure}
\epsfig{file=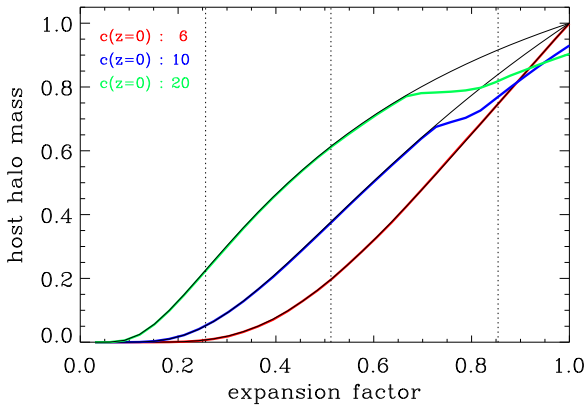,width=0.95\hsize}
\caption{\label{fig:masachi}
Accretion histories for three host haloe concentrations at $z=0$
compared to the analytical description in  Wechsler \& al. (2002),
thin solid lines. The negative deviation from the Wechsler
formula at late times results from the loss of particles in of the
secondary caustic that extends beyond the current virial radius, see
Fig.~\ref{fig:radac}. 
}
\end{figure}
The particles, or later subhaloes, enter the host haloe in proportion
to the dark matter. The growth of the dark matter host haloe is described by
Eq.~\ref{equ:evoma}. The generation of the initial condition for
each particle includes two random selections, the arrival time
and the initial angular momentum. After the first penetration of the
host virial radius the orbits are integrated
numerically. Fig.~\ref{fig:radac} displays the expansion factor at
accretion time versus the current particle distance from the
centre the host, showing distributions for host concentrations of 6, 10
and 20. Subsequently, whenever properties of individual hosts
shall be exemplified we use these tree concentrations
and refer to the accordingly, as the $c_{6}$, $c_{10}$ and $c_{20}$
hosts.  
Despite the angular momentum distribution imposed on
the particles, caustic-like features are apparent as discussed in the
context of the secondary infall model by \cite{Bertschinger85}. 
The most prominent feature is caused by the {\it caustic of the second
turnaround} generated by the particles when they reach the apo-centre 
after their first peri-centre passage. Investigating the dynamics of
the NGC 5846 group \cite{Mahdavi05} find observational support for the
appearance of the caustic of the second turnaround. The location of
the second caustic is correlated with the host concentration. For the
$c_{6}$ host the caustic is located at the virial radius whereas in
the $c_{20}$ host it is found at $\sim1.3$ times the virial
radius. For all host concentrations, the second turnaround is
experienced by particles accreted at $a\sim 0.8$.

Fig.~\ref{fig:masachi} compares the mass growth of the host haloe (thin
solid lines) to the particle accretion history for all particles
located within the host virial radius at $z=0$ (three heavier
lines). Both quantities, host mass and particle numbers, are
normalised to unity. The deviations at late times are caused by  
particles with orbits currently beyond the virial radius also seen in 
Fig.~\ref{fig:radac}.       

The positive correlation between the group-centric distances of the
caustics and the concentrations in Fig.~\ref{fig:radac} arises from
the more rapid mass assembly of high concentration hosts. The thin
solid lines of Fig.~\ref{fig:masachi} indicate that at $a=0.8$. The
$c_{20}$ host has already assembled 90 per cent of 
the final mass compared to only 60 per cent for the $c_{6}$ host. Thus 
at the time ($a\sim0.8$) when the particles are accreted that
presently pass through the second turnaround the $c_{20}$ host
is a factor of 1.5 more massive than the $c_{6}$ host. Therefore
its virial radius is larger by a factor of $1.5^{1/3}=1.14$. However,
Fig.~\ref{fig:radac} shows a factor of $\sim1.3$ difference between
the location of the $c_{20}$ and the $c_{6}$ turnaround. This
additional shrinking of the second turnaround radius
within the less concentrated host is caused by the rapid mass growth of
the $c_{6}$ host at the present time. Thus particles moving outward after
their first peri-centre passage feel a much deeper potential well,
which prevents them from reaching their original starting point, the
first turnaround when they decouple from cosmic expansion. In contrast,
the highly concentrated hosts don't show such a rapid deepening of its 
potential at late times, so the particles can move further out. 
\subsection{Density profiles}
\label{sec:density1}
\begin{figure}
\epsfig{file=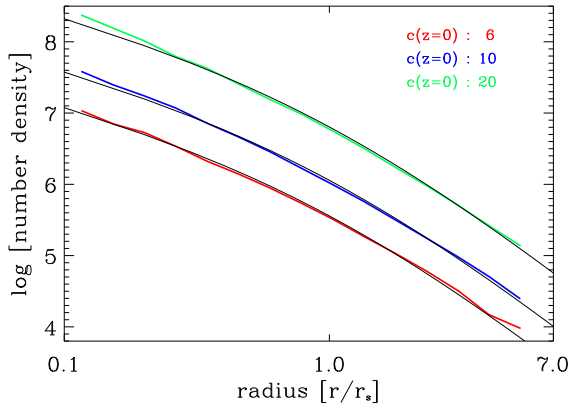,width=0.95\hsize}
\caption{\label{fig:profil}
Arbitrarily scaled number density profiles of the particles for three
host concentrations. The thin solid lines display the NFW-profiles for the
corresponding concentrations. The group-centric distances are scaled
by $r_s$, the characteristic radius of the NFW-profile ($r_s=R\vir/c$).
}
\end{figure}
The host mass grows according to Eq.~\ref{equ:evoma}
and the evolution of the concentration is given by
Eq.~\ref{equ:evoco}. The rate that particles (the same
applies to subhaloes) enter the virial radius of the host is given by
Eq.~\ref{equ:mafu}. About half of all the particles are accreted
onto the host before $z=0.6$. Of course this value varies depending   
on the accretion histories which in our model are determined by the
current concentration of the host. Fig.~\ref{fig:profil} compares the
number density profiles for the three concentrations ($c=6,10,20$)
with NFW-profiles. The agreement of the analytical profiles with
the results of the dynamical evolution is remarkable. The successful
reproduction of the analytical profile by the dynamically evolved
particle distribution provides strong evidence for the capability of
the present model. There appear small deviations at the centre for
the more concentrated hosts which may be caused by a lack of
numerical resolution or by a slight inaccuracy of the employed
accretion formula at early times (see \citealt{Tasitsiomi04,Zhao03}).   
However, this occurs only within a few per cent of the virial radius
and will not affect the results presented here.
\subsection{Velocity dispersion profiles}
\begin{figure}
\epsfig{file=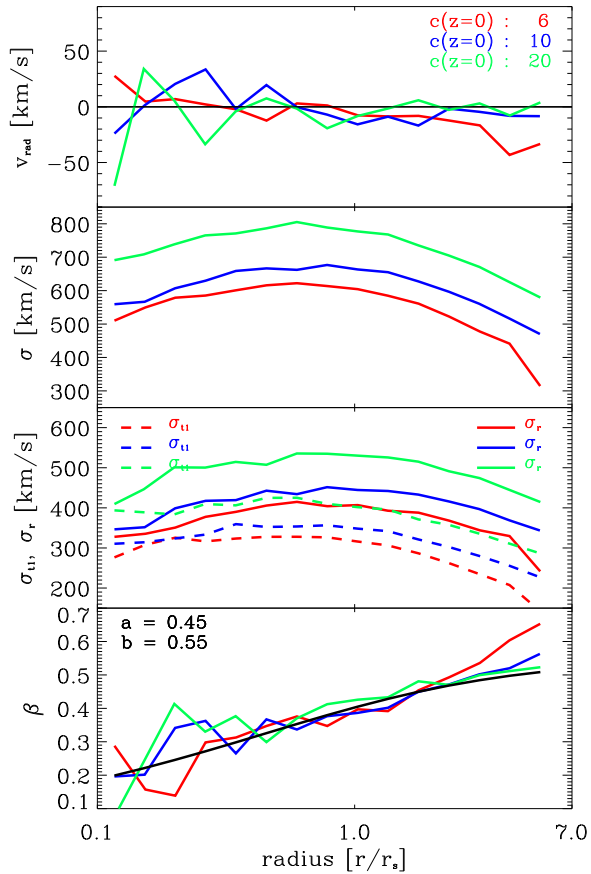,width=0.95\hsize}
\caption{\label{fig:ravel}
The upper three panels display mean radial velocity ($v_{rad}$),
the velocity dispersion ($\sigma$) and the orthogonal velocity dispersion
components (radial $\sigma_r$ and 1D tangential $\sigma_{t,1}$)
of the particles within host haloes of three concentrations. The
group-centric distances are scaled by $r_s$, the characteristic radius
of the NFW-profiles ($r_s=R\vir/c$).  Negative values for the mean
velocities in the fist panel indicate net infall. The lowest panel
shows the associated anisotropy profiles $\beta(r)$. The black solid
line displays the $\beta$-fit given in Eq.\ref{equ:beta}, with an
exponent of $a = 0.45$ and a scaling $b = 0.55$. Note, the anisotropy
profiles of different concentrations can be fitted equally well with
the same parameters.       
}
\end{figure}
We now focus on
the velocity profiles of the point mass sample. In particular we derive
a general expression for the anisotropy parameter
$\beta=1-\left(\sigma_{t,1}/\sigma_{r}\right)^2$ where $\sigma_{r}$ is
the radial velocity dispersion and $\sigma_{t,1}$ is the
uni-dimensional tangential velocity dispersion. If the tangential
velocities are isotropic $\sigma_{t,1} = \sigma_{t} / \sqrt{2}$, where
$\sigma_{t}$ is the two dimensional tangential velocity dispersion.    
By scaling the host-centric distances by $r_s$, the characteristic
radius of the NFW-profiles, one achieves coincidence of the
$\beta$-profiles independent of the concentration.   

The upper panels in Fig.~\ref{fig:ravel} display the radial
mean velocities and the velocity dispersions of the particles
for the three example hosts ($c = 6,10,20$). Negative values in the fist
panel indicate net infall. For the more concentrated hosts
the radial velocities are close to zero, indicating fairly relaxed
systems. The particles within the $c_{6}$ host show some slight
net infall at $\sim 7 r_s$ which my be a result of the recent rapid
growth phase of the host haloe. The velocity dispersions in the second
panel show a behaviour known from N-body simulations \citep{Hoeft04},
they peak at about $0.8 r_s$. In the third panel the velocity
dispersions are split up into radial and tangential components. To
allow for a direct comparison with the radial component the
uni-dimensional tangential velocity dispersion $\sigma_{t,1}$ is shown
here. Finally, the anisotropy profiles $\beta(r/r_s)$ for the three host
haloes are given in the lowest panel. Despite the
different host concentrations, we find that the anisotropy profiles
are very similar if the distances scaled by the characteristic radius
$r_s$ of the NFW-profile rather than the virial radius $R\vir$.
The $\beta$-profiles can be fit with an 
concentration-independent expression:  
\begin{equation}
\label{equ:beta}
\beta(r) = b \left({x\over x+1}\right)^a\hspace{0.2\hsize}x=r/r_s
\end{equation}
This expression enables an analytical integration of
the Jeans equations as discussed in the Appendix. 

In addition to the successful reproduction of the density profiles the
present inspection of the velocity profiles confirms that the dynamical
properties of our model are also in good agreement with the
dynamical characteristics of CDM haloes generated by N-body
experiments. Based on 10 relaxed dark matter halos \cite{Wojtak05}
find a similar shape and asymptotic values for the average
$\beta$-profile. Also the $\beta$-profiles based on six highly resolved
N-body systems presented by \cite{Dehnen05} (based on simulations
discussed in \cite{Diemand04a} and \cite{Diemand04b}) show very
similar features. Finally, the density slope - velocity anisotropy
relation discussed in \cite{Hansen06} implies a self similarity of
$\beta$-profiles when scaled by $r_s$ as advocated here.
\subsection{Concentration-sigma relation}
\begin{figure}
\epsfig{file=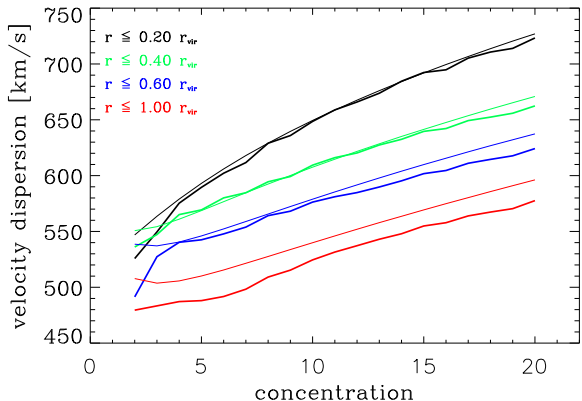,width=0.95\hsize}
\caption{\label{fig:fvir}
Mean velocity dispersions within the listed fractions of the virial
radius versus the host concentration. The thick lines display the
results from the dynamical models and the thin solid lines arise from
the numerical integration of the Jeans equation using the 
$\beta$-formula Eq.~\ref{equ:beta} (see Appendix).    
}
\end{figure}
We now investigate the relation between the mean particle
velocity dispersion and the concentration of the host haloes. It is
expected that the velocity dispersion increases with concentration,
therefore the present results should be considered as an
further affirmation of the model and a point of reference for
subsequent investigations of the subhaloes sample. 

Fig.~\ref{fig:fvir} shows the dependence of the mean particle velocity
dispersions on the host concentrations. The mean values of the
dispersions are computed within several fractions of the virial
radius. The thin solid lines display the results from the analytical
integration of the Jeans equation using Eq.~\ref{equ:beta} to describe
the radial behaviour of the anisotropy parameter. Details of this
integration can be found in the Appendix. The agreement between the
model and the analytical computation is convincing. For
very small concentrations $\lesssim 5$ some deviations appear,
but hosts haloes with these small concentrations are in the process of
formation and the applicability of the Jeans equation in this form may
be questioned. The $\sim5$ per cent deviation for the mean dispersion
within the full virial radius $1.00 R\vir$ probably results from the 
orderly accretion mode inherent in our model. 
Its traces are certainly not blurred out before the second turnaround
passage (see also Fig.~\ref{fig:radac}). The main finding is that the
velocity dispersion for hosts with concentrations $\sim20$
are a factor of $1.2 - 1.3$ higher then the mean velocity dispersions
of equal mass hosts with concentrations $\sim5$. The mean central
velocity dispersions increase somewhat stronger with the host
concentration than those within larger radii.  

In summary, our comprehensive examinations of the point mass sample
have shown satisfactory agreement with the results from N-body
simulations, indicating that our simplified dynamical model can
reproduce the main characteristics found by more elaborate N-body
simulations.  However, we notice that the outskirts ($\gtrsim
0.6R\vir$) may be slightly affected  by the unphysical regularity of
the accretion mode. Nevertheless, the model enables us to measure in
detail the dependence of the subhaloe properties on the host
concentration. 
\section{Results for the subhaloe samples}	
\label{sec:subhaloe}
This section describes the dynamical evolution of extended subhaloes
within the growing potentials of their hosts. As described in
\S~\ref{sec:subprop} we attribute to every (sub)haloe a NFW-density
profile and include the effects of dynamical friction and tidal
truncation in the integration of the orbits. For every host we
distinguish two subhaloe samples at $z=0$, a {\it complete} sample and
a {\it tidally reduced} sample. The former comprises all 100000
subhaloes independently of their present truncation radius and the
latter counts only those subhaloes whose tidal truncation radius is
larger then $3\kpc$. The tidally reduced sample is a sub-sample of the
complete sample. For example, the reduced samples of the $c_{6}$,
$c_{10}$ and $c_{20}$ host haloes comprise a number of 74777, 57218,
36258 surviving subhaloes, respectively, of the original 100000 within the complete
sample. The idea behind this distinction is to imitate the effects of
detection limits which are inherent to observation and simulations. If
satellite galaxies fall below a certain luminosity, they won't appear
in the catalogue. And if subhaloes fall below a certain particle limit
they won't be counted as substructure. In both cases the disappearance
from the catalogue does not necessarily mean total physical
disassembly. The accretion times and the assembly history of the
subhaloes samples do not differ from those plotted in
Figs.~\ref{fig:radac} and \ref{fig:masachi}) for the point mass
sample. Consequently, we start the investigation of the subhaloe
samples with the discussion of the evolution of the mass function. 
\subsection{Mass functions}
\begin{figure}
\epsfig{file=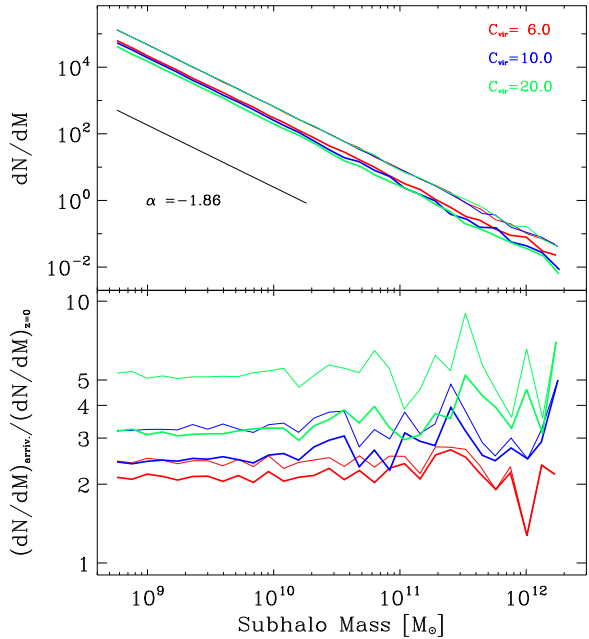,width=0.95\hsize}
\caption{\label{fig:maha}
The upper panel compares the differential mass function of the
subhaloes in the complete sample before entering the virial radius of
the host (thin lines) with the mass function of the subhaloes at $z=0$
(thick lines). The straight line displays the slope used in
Eq.~\ref{equ:mafu}. There is a week dependence of the amplitude of the
final mass function on the  concentration. The lower panel shows the
ratios of the initial to the present mass functions. Thick lines are
used for the complete subhaloe sample (these values can be directly
read off the upper panel). The thin lines in the lower panel display the
ratios between the mass function of the initial subhaloe mass function
to the tidally reduced sample.
}
\end{figure}
\begin{figure}
\epsfig{file=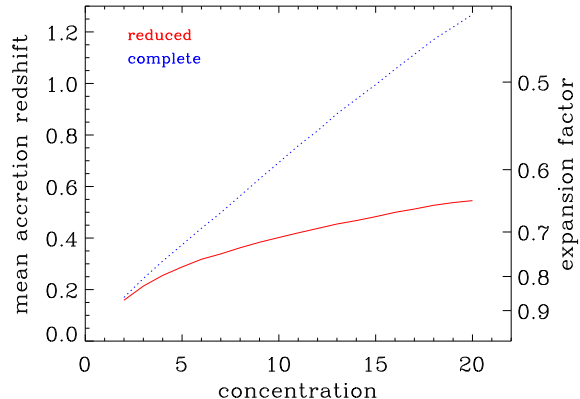,width=0.95\hsize}
\caption{\label{fig:mzac}
Variation of mean accretion redshift of subhaloes with host
concentration at $z=0$. The dotted line shows the results for the
complete subhaloe sample. The solid line displays the relation for the
tidally reduced subhaloe sample.} 
\end{figure}
The assignment of initial subhaloe masses is designed to reproduce 
a differential mass distribution $dN/dM$ with a power law shape 
and a logarithmic slope of $\alpha = -1.86$, the appropriate
slope for field haloes measured in N-body simulations. The upper panel
of Fig.~\ref{fig:maha} compares the differential mass functions of the
arriving subhaloes (thin lines) with the mass function of the
complete subhaloes at $z=0$ after they were dynamically processed
within the three example hosts ($c=6$, $10$, $20$).  
The mass functions are shown within a range of 
$5\times 10^8-2\times10^{12}\Msol$. The initial mass
functions for the three concentrations perfectly coincide, indicating
that the temporarily slightly different upper mass limits due to the
altering host growth rates (see Eq.~\ref{equ:mafu}) do not affect the
overall mass function. Interestingly the logarithmic slope does not
change for the dynamically processed subhaloe samples at
$z=0$. Similar results  have been found in N-body simulations
(e.g. \citealt{Reed05}). There appears a slight dependence of the
amplitude of the dynamically processed mass functions on the host
concentrations. More concentrated hosts have slightly reduced mass
functions. The reason for the weak dependence on host concentration
can be explained by the correlation between mean subhaloe accretion
times and host concentrations as displayed in
Fig.~\ref{fig:mzac}. On average subhaloes in more concentrated hosts
are accreted earlier, thus they are exposed to the tidal field of the
host for a longer time and they suffer more severe mass
losses. The thick lines in the lower panel of Fig.~\ref{fig:maha}
may clarify this issue. They display the ratios of initial to the
final subhaloe mass function. The $c_{6}$ host exhibits a difference
in amplitude of a factor of $\sim2$, whereas in the $c_{20}$ host
the final mass function is reduced by a factor of $\sim3$. The thin
lines in the lower panel display the ratios between the initial and
the final mass distributions in the tidally reduced subhaloe
sample. At the low mass end the difference between initial and final
mass function is enhanced, however the high mass end shows barely any
difference. Therefore, at the low 
mass end the power law slope is conserved even in the tidally reduced
sample. The solid line in Fig.~\ref{fig:mzac} displays the mean
accretion times as a function of host concentration for the tidally
reduced sample. In particular for concentrations $\sim20$ the
difference of the mean accretion time is about $3.5\Gyr$.  
\subsection{Density profiles}
\begin{figure}
\epsfig{file=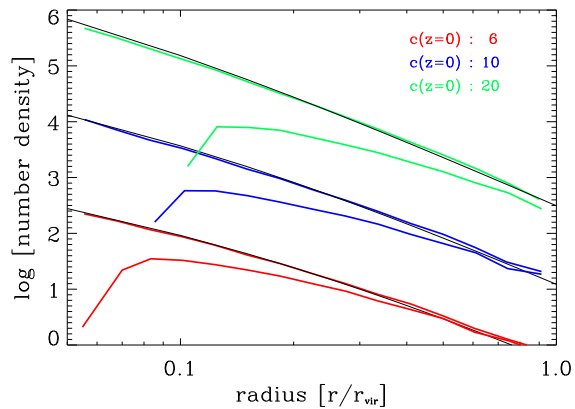,width=0.95\hsize}
\caption{\label{fig:provir}
Arbitrarily scaled number density profiles for the complete and
the tidally reduced subhaloe sample within host haloes of three
concentrations (c=6,10,20). The thin solid lines give the associated
NFW-profiles. The complete sample shows perfect agreement with
the NFW-profiles. The profiles of the tidally reduced samples 
flatten towards the centre. Subhaloe populations in more
concentrated hosts are more efficiently reduced. 
}
\end{figure}
Fig.~\ref{fig:provir} shows the number density profiles for the
complete and  the tidally reduced subhaloe samples within the
three host haloes with present concentrations of 6, 10 and 20.
In contrast to the density profiles in Fig.~\ref{fig:profil} the
distances from the centre of the host are scaled by the virial radius
$R\vir$.  
For comparison the associated NFW-profiles are displayed with thin
solid lines. Similar to findings for the point mass sample in
\S~\ref{sec:density1}, the agreement between the NFW-profile and the
complete sample is remarkable, indicating that the combined
action of dynamical friction and tidal truncation does not severely
alter the number density profile. Subhaloes lose a substantial fraction of
their mass during their first peri-centre passage, after that
the dynamical friction is strongly reduced since dynamical friction
drag is proportional to the subhaloe mass (Eq.~\ref{equ:chan}). The
number densities of all tidally reduced samples show a centrally
flattened behaviour, indeed at the very centre they display positive
gradients. We do not discuss to what extent these positive gradients
are physical since the presence of 
baryons in subhaloes certainly alters their tidal
resistance. However, it is noteworthy that most of tidal reduction
within the $c_{6}$ host is confined to the volume within half the virial
radius whereas the tidal reduction is visible over the whole haloe
volume in the case of the $c_{20}$ host.
\subsection{Velocity dispersion profiles}
\begin{figure}
\epsfig{file=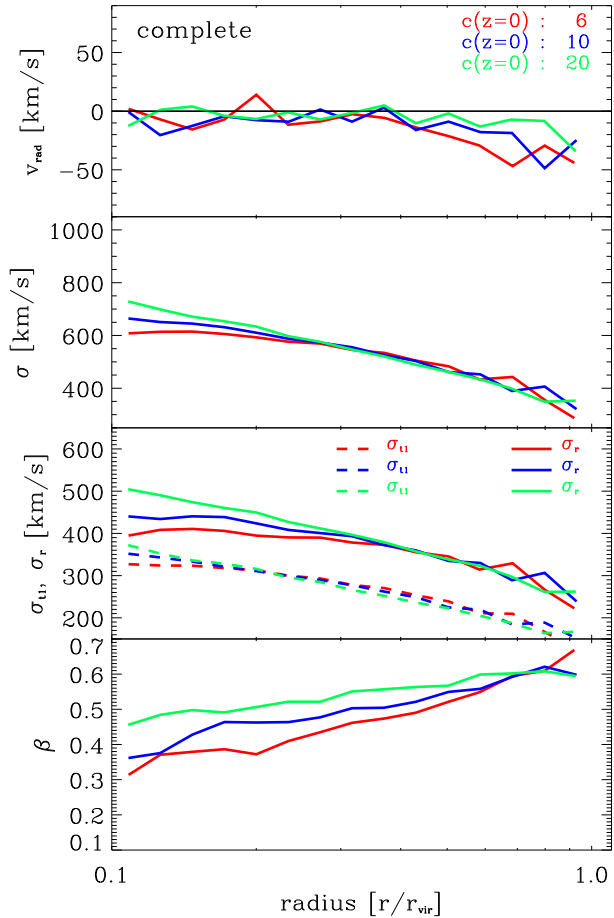,width=1.00\hsize}
\caption{\label{fig:ravel_DFTRM}
The upper three panels display mean radial velocity ($v_{rad}$),
the velocity dispersion ($\sigma$) and the orthogonal velocity dispersion
components (radial $\sigma_r$ and 1D tangential $\sigma_{t,1}$) of the
complete subhaloe population within host haloes of the indicated
concentrations. As distinct from Fig.~\ref{fig:ravel}, here the
group-centric distances are scaled by virial radius $R\vir$. 
}
\end{figure}
\begin{figure}
\epsfig{file=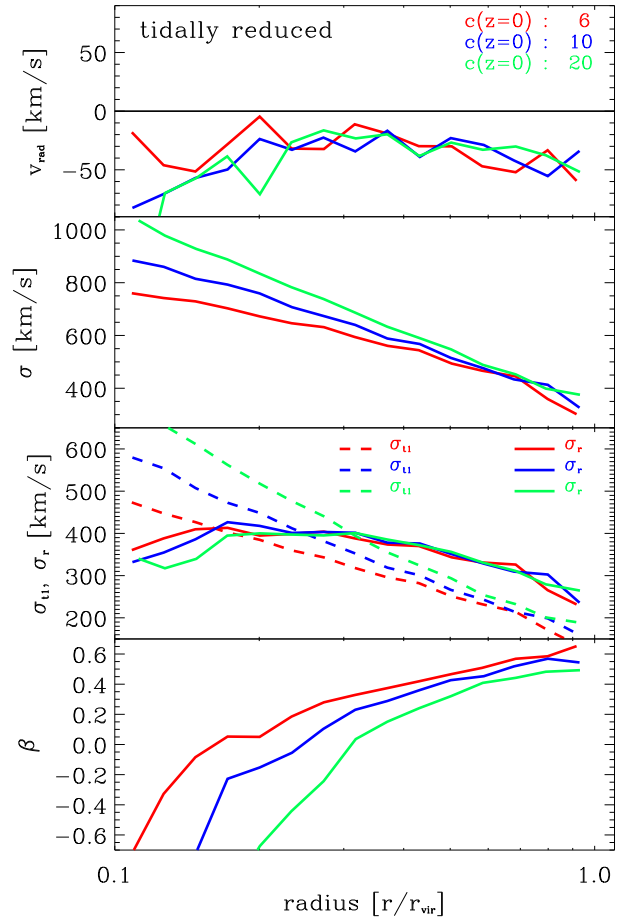,width=1.00\hsize}
\caption{\label{fig:rupel_DFTRM}
The upper three panels display mean radial velocity ($v_{rad}$),
the velocity dispersion ($\sigma$) and the orthogonal velocity dispersion
components (radial $\sigma_r$ and 1D tangential $\sigma_{t,1}$) of the
tidally reduced subhaloe population within host haloes of the indicated
concentrations. As distinct from Fig.~\ref{fig:ravel}, here the
group-centric distances are scaled by virial radius $R\vir$. 
}
\end{figure}
Some very interesting insights into the dynamics of subhaloes can be
gained by comparisons of the velocity distributions for the
complete and the tidally reduced samples displayed in
Figs.~\ref{fig:ravel_DFTRM} and \ref{fig:rupel_DFTRM} respectively.
It is important to keep in mind that the tidally reduced sample is a
subsample of the complete sample. Therefore, higher mean velocities
in the reduced sample can only be achieved if preferentially slow
moving haloes are removed. In other words there is no source of high
velocity haloes in the reduced sample. The uppermost panel in
Fig.~\ref{fig:ravel_DFTRM} displays the mean
radial velocity of the complete sample. With exception of the
outskirts they are close to zero, 
indicating a relaxed state. On the other side the mean velocities
of the tidally reduced samples (Fig.~\ref{fig:rupel_DFTRM}) show a
clear trend for inward motion, 
negative $v_{rad}$. This is most naturally explained by 
tidal shrinkage as subhaloes first approach the host centre. 
Consequently, most probably the truncation radii 
falls below the rejection radius of $3\kpc$ close to the peri-centre
passage of the subhaloe, resulting in a lack of outward streaming
subhaloes in the reduced sample. For large radii the velocity
dispersions  of the two samples are similar (second panels in
Figs.~\ref{fig:ravel_DFTRM} and \ref{fig:rupel_DFTRM}). However, the
central velocity dispersions of the reduced sample are
$\sim1.3$ times larger than in the complete sample. Again this
increase of the velocity dispersion can only be explained by the lack
of preferentially slow moving subhaloes. The comparison of the
orthogonal components of the velocity dispersions (third panel in
Figs.~\ref{fig:ravel_DFTRM} and \ref{fig:rupel_DFTRM}) indicates that
the increase of the central velocity dispersion in the reduced sample is
based on the strong enhancement of the tangential dispersions, which
again must arise due to a lack of slow tangential motions. The
radial components give the reverse picture. The central radial
velocity dispersion of the reduced sample is actually smaller,
indicating a deficit of fast radially moving subhaloes compared to the
complete sample. A hint for this mechanism was reported by
\cite{Faltenbacher05a} investigating the velocity distribution of
galaxies in hydro-dynamical simulations of clusters of galaxies.          
The resulting anisotropy profiles are quite different (note the
different scaling in the lowest panels of Figs.~\ref{fig:ravel_DFTRM}
and \ref{fig:rupel_DFTRM}). If one assumes similar processes operating 
for satellite galaxy populations in simulated groups, the negative
values for $\beta$ displayed in \cite{Benatov06} (their Fig. 5) are in
agreement with the findings presented here. \cite{Maccio06} discussed
the enhanced survival rate of subhaloes with condensed baryonic
cores. In theory the behaviour of the velocity distribution of
satellite galaxies in groups can be used to determine their tidal 
disruption rate.
\subsection{Concentration-sigma relation}
\begin{figure}
\epsfig{file=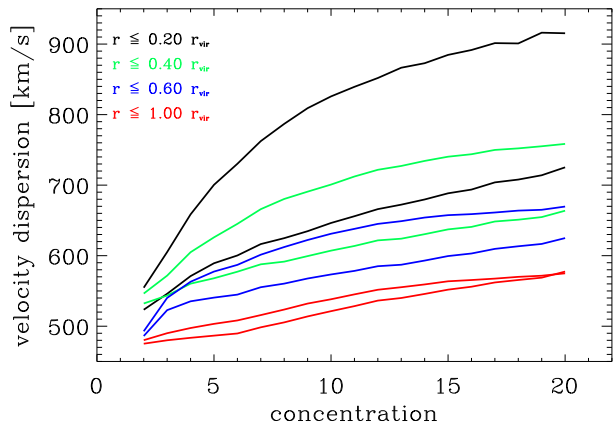,width=1.00\hsize}
\caption{\label{fig:rup2}
Variation of mean subhaloe velocity dispersions with 
concentration of the host for various inclusion radii as listed. 
The upper line of each pair of lines with the same line style
always displays the dispersion of the tidally reduced sample.
The lower lines show the values belonging to the complete sample.
}
\end{figure}
In analogy to Fig.\ref{fig:fvir} we display in Fig.\ref{fig:rup2} the
mean dispersions within 0.2, 0.4, 
0.6 and 1 times the virial radius $R\vir$ for the complete and
the tidally reduced subhaloe samples. The larger dispersions always
correspond to the tidally reduced sample whereas the lower line of
each pair refers to the complete sample. The dispersions of the 
point mass sample in Fig.\ref{fig:fvir} are quite similar to the
values of the complete sample, indicating only a weak
impact of dynamical friction on the global properties of the subhaloe
population. However, the impact of tidal reduction is enormous for the
central dispersions, e.g. for host concentrations of $\sim20$
the velocity dispersions within $0.2 R\vir$ are enhanced by a factor
of $1.3$ compared to the complete sample. The dependence on
concentration within the tidally reduced sample is also more
prominent, the central dispersion in hosts with concentration
$\sim5$ to hosts with concentration $\sim20$ increases by a factor
$1.4$. The mean dispersion within the virial radius is
only weakly affected by the tidal reduction. 

Since the complete sample resembles the behaviour of the point mass
sample, the Jeans equation holds also in this case. 
Due to the negative mean radial velocities in the reduced sample
(Fig.~\ref{fig:rup2}) one of the preconditions for the description by 
the Jeans equation is no longer valid (see Appendix). We do not aim
to present a modified solution which can cope with radial motions, 
because the extent of inflow depends on the choice of the rejection
radius in our model. 

The important result here is that an increase in central velocity
dispersion of satellite galaxies in groups caused by higher
concentration may be amplified if some fraction of the satellite
population has been  tidally disrupted. Both trends are in the same
direction. Low concentration hosts have relatively 
low central velocity dispersions and their recent formation times make
tidal disruption as a driver for an enhanced velocity dispersions
unlikely. Strongly concentrated hosts have higher velocity
dispersions and accreted most of their satellites early on. This
old population may be more severely reduced by tidal forces, causing
an additional boost to the central velocity dispersion. 
\section{Summary}	
\label{sec:summary}
Using a dynamical model for the evolution of subhaloes within a
growing host potential in galaxy groups, we have investigated the
relation between host concentration and properties of subhaloe
population at $z=0$. The concentration is correlated with the group
formation redshift. Groups with 
equal virial masses at $z=0$ but different formation times show
significant differences in their subhaloe properties. As a base point we
have computed the dynamical evolution of point mass samples and their
dependence on the host concentration. The model then has been adapted
to cope with the spatial extent of subhaloes and the impact of dynamical
friction and tidal disruption. We only focus on the dynamical
evolution of subhaloes that are massive enough to possibly host
galaxies. Every substructure analysis is subject to detection limits
whether it is the sensitivity limit in connection with observations or
the resolution limit in numerical simulations. To study the effects of
detection limits, we create a {\it reduced} subhaloe sample comprising
only those subhaloes with current radii larger than a rejection radius of
$r\rej = 3\kpc$. The reduced sample is a sub-sample of the {\it complete}
sample, comprising all subhaloes independent of the actual truncation
radius. The rejection radius can be considered as a parameter of the 
model and has been calibrated by observations as well as numerical
simulations (see \citealt{Faltenbacher05}). Originally the model
is set up to trace the evolution of the dark matter subhaloes in a
group environment, however if the subhaloes host luminous galaxies, a
tidal truncation below $3\kpc$ most probably would affect the stellar
component of the galaxy as well.   

Our main findings are: 
(1) The velocity dispersion of subhaloes or satellite galaxies in
groups depends on the concentration of the underlying dark 
matter distribution of the host haloe. Equal mass hosts with higher
concentrations have enhanced  velocity dispersions of the
subhaloes, in particular at the centre. Since higher
host concentrations are caused by earlier formation times, this
relation implies that subhaloe populations with early formed hosts
exhibit higher velocity dispersions than those residing in
more recently formed hosts of equal mass.
(2) We propose a fitting formula for the radial dependence of the
anisotropy parameter $\beta$, which is independent of concentration 
if the group-centric distances when scaled by the characteristic radius
of the NFW-profile $r_s$.  
(3) The increase of the central velocity dispersion with concentration
is amplified in the reduced sample. This is caused
by an enhancement of the tangential velocity dispersion and a lack of
slow radial motions compared to the complete sample.  
(4) The dynamical evolution of subhaloes that experience
tidal stripping and dynamical friction does not alter the slope of the
mass function, in agreement with the results from N-body
simulations (see e.g. \citealt{Reed05}), but dynamical evolution
reduces the amplitude by a factor of $\sim2$. In the complete sample
the amplitude of the mass function is only weakly  dependent upon the host
concentration. However, the reduced sample shows a stronger dependence
on host concentration with amplitudes decreasing with increasing host
concentrations. For a host concentration of 20 the amplitude of the
initial mass function decreases by a factor of $\sim5$, however this
value depends strongly on the choice of the rejection radius.

The present investigation reveals the difficulties associated with
mass estimates derived from the velocity dispersion of satellites in
groups. The concentration of the host haloe and the effects of tidal
reduction of the satellite galaxies in groups may alter the central
velocity dispersions. The appearance of intra-group stars is strong
evidence for the reduction of the satellite luminosities.   
According to \cite{Osmond04} the velocity dispersion provides a very
unreliable measure of system mass. The concentration and its influence
on tidal reduction add more concerns on the accuracy of such
mass estimates. 

Combined observations of X-ray temperatures and velocity dispersions
of satellite galaxies have recently become available for groups. As a 
application of the present analysis these observations my be used to
infer the impact of tidal forces onto the central satellite
population. If tidal forces in fossil groups efficiently reduce the
number of detectable satellites, we expect the central velocity
dispersion to be high compared to the X-ray temperature. Some
support for this scenario has been found by \cite{Khosroshahi06}. 

A comparison of the X-ray temperatures of the 
hot intra-group gas with the velocity dispersions found for the
satellite galaxies can be used to infer the degree of tidal
disruption. A flattening of the central number density profiles along
with an increasing velocity dispersion are the signatures of tidal
reduction of the central subhaloe or satellite population.  

\cite{SommerLarsen06} considers radial velocity dispersions in
fossil systems at a fixed radius $r\gtrsim30\kpc$. We have presented 
a uniform fitting formula for the anisotropy parameter provided the
group-centric distances are scaled by $r_s$, the characteristic radius
of the NFW-profile of the host. Fossil groups are assumed to be early
formed systems (see e.g.~\citealt{DOnghia05}) with high
concentrations. Our fitting formula implies that at a fixed radius
more concentrated hosts generate more radially anisotropic velocity
dispersions. This result has to be taken into account if the velocity
dispersions of different groups are compared at physically similar
radii.
\section*{Acknowledgements}
The authors would like to thank the anonymous referee for the
constructive comments which helped to improve the text. Useful
discussions with Juerg Diemand and Stelios Kazantzidis are highly
appreciated. This work has been supported by NSF grant AST 
00-98351 and NASA grant NAG5-13275 for which we are very grateful.

\section*{Appendix: Integration of the Jeans Equation}
\label{sec:appendix}
Fig.~\ref{fig:fvir} compares the velocity dispersions from our
dynamical model with the predictions of the Jeans equation. For a 
spherical symmetric and static ($\bar{\sigma}_r = \bar{\sigma}_t = 
0$, no mean radial or tangential motions) system the Jeans equation
can be written as 
\begin{equation}
\label{equ:jeans}
{1\over\rho}{\rm{d}\over\rm{d}r}(\rho\sigma^2_r)
+
2\beta{\sigma^2_r\over r} 
= 
-{\rm{d}\Phi\over\rm{d}r}
\end{equation}
where $\rho$ is the density, $\sigma_r$ is the radial velocity
dispersion, $\Phi$ is the gravitational potential and $\beta$ is the
anisotropy parameter (see e.g.~\citealt{Binney87}). A solution of this
first order differential equation can be be obtained by means of
multiplying both sides of the Eq.~\ref{equ:jeans} by the integrating
factor   
\begin{equation}
\label{equ:intfac}
u(r) = \ \exp\left[ 2\int_0^r{\beta(r^\prime) d r^\prime \over
r^\prime}\right]
\end{equation}
and performing an integration from a given radius $r$ to infinity.  
The application of the boundary condition $\lim_{r\rightarrow
\infty}\sigma^2_r=0$ which any ordinary bound system satisfies results
in        
\begin{multline}
\rho(r)\sigma_r^2(r)\ \exp\left[2\int_0^{r}{\beta(r^{\prime}) d r^{\prime}
\over r^{\prime}}\right] 
\\=
\int_r^\infty {\rm{d}\Phi\over\rm{d}r}(r^\prime)\rho(r^\prime)
\ \exp\left[2\int_0^{r^\prime}{\beta(r^{\prime\prime}) d r^{\prime\prime}
\over r^{\prime\prime}}\right] dr^{\prime}\ .
\end{multline}
For completeness we repeat the fitting formula for the anisotropy
parameter $\beta$ as given in Eq.~\ref{equ:beta}   
\begin{equation}
\label{equ:betaapendix}
\beta(r) = b \left({x\over x+1}\right)^a\hspace{0.2\hsize}x=r/r_s\ ,
\end{equation}
where $r_s$ is the scale factor of the NFW-profile. With this
expression for $\beta(r)$ the integrating factor
(Eq.~\ref{equ:intfac}) can be written as    
\begin{equation}
\hat{u}(x) = u(xr_s) = \exp\left\{2b\ _2F_1[a,a,1+a,-x]\right\}
\end{equation}
where $_2F_1$ is the hyper-geometric function and $x = r/r_s$
as introduced in the foregoing equation. Finally, the mass weighted
mean velocity dispersion within the radius $r$ is given as  
\begin{equation}
\bar{\sigma}(r) = {
\int_0^r\rho(r^\prime)\sigma_r^2(r^\prime)r^{\prime2}dr^\prime
\over
\int_0^r\rho(r^\prime)r^{\prime2}dr^\prime\ .
}
\end{equation}  
The mean velocity dispersions represented by the thin lines in
Fig.~\ref{fig:fvir} are computed on the assumption that the 
matter distribution follows a NFW-density profile with a virial mass
of $3.9\times10^{13}\Msol$. As seen in Fig.~\ref{fig:radac} up to 10
per cent of all the particles are located outside the virial radius
at $z=0$. This particle deficit at large radii may explain the
$\sim5$ per cent deviations between the velocity dispersions obtained
from the model and the Jeans equation ($r\leq1.00R\vir$ line in
Fig.~\ref{fig:fvir}).  
\end{document}